\documentclass[reprint,
superscriptaddress,
amsmath,amssymb,aps,showkeys,showpacs,
twoside,final,secnumarabic,%raggedbottom,
nofootinbib]{revtex4-2}

\usepackage[paperwidth=205mm,paperheight=290mm,top=17mm,bottom=25mm,
inner=17mm,outer=17mm,
twoside]{geometry}

\usepackage{cmap} % Улучшенный поиск %русских слов в~полученном pdf-файле
\usepackage[T1,T2A]{fontenc}
\usepackage[utf8]{inputenc}
\usepackage[russian,english]{babel}
\usepackage{color}
\usepackage{graphicx}% Include figure files
\usepackage{dcolumn}% Align table columns on decimal point
\usepackage{bm} % bold math
\usepackage[unicode=true,colorlinks=true,linkcolor=magenta, urlcolor=blue, citecolor = blue,breaklinks]{hyperref}
\usepackage{multirow}
\usepackage{url}
\usepackage{breakurl}
\usepackage{float}
\usepackage{natbib}
\usepackage[framemethod=TikZ]{mdframed}
\DeclareGraphicsExtensions{.eps}

\newcount\issue
\newcount\Vol
\newcount\numb
\usepackage{fancyhdr} %this packages %provides fancy up and bottom of page
\def\Vol{\textbf{79}}
\def\numb{x}
\begin{document}

%====== Начало шапки статьи  ============
\title{MACHINE LEARNING IN NATURAL SCIENCES
 \\[20pt] Engineering Point Defects in MoS2 for Tailored Material \\ Properties using Large Language Models}

\def\addressa{HSE University, Myasnitskaya Ulitsa, 20, Moscow, Russia, 101000}
\def\addressb{Institute for Functional Intelligent Materials, National University of Singapore, 4 Science Drive 2, Singapore 117544}
\def\addressc{Constructor University Bremen, Campus Ring 1, Bremen, 28759, Germany}

\author{\firstname{Abdalaziz}~\surname{Al-Maeeni}}
\email[E-mail: ]{al-maeeni@hse.ru}
\affiliation{\addressa}

\author{\firstname{Denis}~\surname{Derkach}}
\affiliation{\addressa}

\author{\firstname{Andrey}~\surname{Ustyuzhanin}}
\affiliation{\addressb}
\affiliation{\addressc}

\received{xx.10.2024}
\revised{xx.xx.2024}
\accepted{xx.xx.2024}

\begin{abstract}
The tunability of physical properties in transition metal
dichalcogenides (TMDCs) through point defect engineering offers
significant potential for the development of next-generation
optoelectronic and high-tech applications. Building upon prior work on
machine learning-driven material design, this study focuses on the
systematic introduction and manipulation of point defects in MoS2 to
tailor their properties. Leveraging a comprehensive dataset generated via density functional theory (DFT) calculations, we explore the effects of
various defect types and concentrations on the material
characteristics of TMDCs. Our methodology integrates the use of
pre-trained large language models to generate defect configurations,
enabling efficient predictions of defect-induced property
modifications. This research differs from traditional methods of
material generation and discovery by utilizing the latest advances in
transformer model architecture, which have proven to be efficient and accurate discrete predictors. In contrast to high-throughput methods where
configurations are generated randomly and then screened based on their
physical properties, our approach not only enhances the understanding
of defect-property relationships in TMDCs but also provides a robust
framework for designing materials with bespoke properties. This
facilitates the advancement of materials science and technology.
\end{abstract}


\keywords{TMDC, MoS2, defects, transformers, Llama   \\[5pt]}

%DOI:  

\maketitle
\thispagestyle{fancy}

%====== Начало  статьи  ============

\section{Introduction}\label{intro}
Transition metal dichalcogenides (TMDCs) are a promising class of materials that have garnered significant attention in recent years due to their unique structural and electronic properties \cite{manzeli20172d,wang2012electronics}. Composed of transition metals such as molybdenum (Mo) and tungsten (W), and chalcogen elements like sulfur (S), selenium (Se), and tellurium (Te), TMDCs possess a layered structure analogous to that of graphene \cite{wang2012electronics}. This structure is characterized by strong covalent bonds within the layers and weak van der Waals forces between them, facilitating exfoliation into monolayers and enabling diverse applications.

TMDCs encompass a wide spectrum of properties from metallic to semiconducting behaviors. This versatility makes them a focal point for research in electronic and optoelectronic devices. TMDCs can introduce localized states in the bandgap, which significantly affect conductivity and carrier mobility. Such modifications are critical for optimizing the performance of electronic devices and tailoring their properties for specific applications.

The optical properties of TMDCs are equally significant, influencing photoluminescence and absorption spectra. These properties are crucial for optoelectronic applications such as photodetectors, solar cells, and light-emitting devices. TMDCs exhibit strong light-matter interactions, positioning them as promising candidates for these technologies. Furthermore, their bandgap can be tuned by altering their thickness, opening avenues for tailored optoelectronic properties.

In addition to their electronic and optical versatility, TMDCs exhibit high mechanical flexibility and strength, which are essential attributes for the development of flexible and wearable electronic devices. Point defects within TMDCs can affect their mechanical properties, influencing both the strength and flexibility of the material. These mechanical properties, combined with their electronic and optical characteristics, underscore the potential of TMDCs in next-generation flexible electronics and photonics.

The engineering of materials with precise properties is a central goal in contemporary materials science, and the study of two-dimensional (2D) materials, such as transition metal dichalcogenides (TMDCs), provides unique opportunities to achieve this. TMDCs exhibit a remarkable susceptibility to modifications through chemical alterations, particularly through the introduction of defects. These defects, which can be incorporated as adatoms on surfaces or as substitutions within the crystal lattice, play a crucial role in tailoring material properties and broadening their application potential.

In TMDCs, defects significantly influence electronic properties by introducing localized states within the bandgap. These states can affect charge transport by serving as traps or scattering centers, thereby altering conductivity and carrier mobility. Such modifications are essential for optimizing TMDCs in electronic applications, such as transistors and sensors, where specific electronic characteristics are required.

Defects also have a profound impact on the optical properties of TMDCs. By affecting exciton recombination pathways, defects can alter photoluminescence and absorption spectra, which is vital for optoelectronic applications, including photodetectors, solar cells, and light-emitting devices. The ability to precisely control defect types and concentrations enables the tuning of optical responses, enhancing efficiency and functionality in these technologies.

Moreover, defects influence the mechanical properties of TMDCs, affecting their strength and flexibility. Defects can serve as stress concentration sites or facilitate deformation, impacting the material's mechanical integrity and pliability. Engineering defects is crucial for the development of flexible electronics, where materials must maintain robustness while enduring bending and stretching.

Despite the potential of defect engineering in TMDCs, predicting the effects of defects remains challenging due to the vast configurational space and complex quantum mechanical interactions involved. Advances in computational methods, such as high-throughput simulations and machine learning, are being utilized to tackle these challenges. However, the prediction of defect properties using machine learning is limited by the availability of comprehensive datasets and the complexity of quantum state predictions.

In order to engineer crystals with specific physical attributes, a comprehensive dataset covering the defect configuration space is provided by our previous research \cite{huang2023unveiling}. This dataset enables creation of more accurate machine learning models, facilitating the exploration and optimization of material properties for emerging technologies. In this work, we propose a new application of an open source pre-trained large language model to generate defects configuration conditioned on physical properties. We show promising results in the domain of defects engineering and possibly even full crystal structure generation. We hope this research will set the scene for further LLMs applications in crystal defects generation.

\subsection{Purpose and Objective}
The primary purpose of this work is to advance the understanding and engineering of transition metal dichalcogenides (TMDCs) by leveraging computational methods to predict and tailor material properties through defect engineering. This research aims to address the growing demand for materials with specific electronic, optical, and mechanical characteristics by providing a systematic approach to designing TMDCs with tailored defects that meet predefined property criteria. The objective of this study is to develop a framework that enables the generation of crystal structures with defects that satisfy specific target properties. These properties include key electronic and physical attributes such as energy levels, the highest occupied molecular orbital (HOMO), the lowest unoccupied molecular orbital (LUMO), bandgap, formation energy, and other relevant physical characteristics. By achieving this objective, the work seeks to facilitate the design of TMDCs for targeted applications in electronics, optoelectronics, and flexible devices.

The model uses several inputs to guide the design of TMDCs with desired defect configurations. These inputs include energy levels, which influence the stability and reactivity of the crystal structure. The highest occupied and lowest unoccupied molecular orbitals (HOMO and LUMO) are crucial for understanding the electronic properties of the material, such as conductivity and reactivity. The bandgap, defined as the energy difference between the HOMO and LUMO, determines the material's semiconducting properties and its suitability for electronic and optoelectronic applications. Formation energy, the energy required to form a defect within the crystal lattice, is essential for assessing the feasibility and stability of defect configurations. Other physical attributes, such as mechanical strength, flexibility, and optical characteristics, may also influence the material's performance.

The core objective of this work is to utilize these inputs to generate crystal structures with defects that satisfy the specified properties. By developing and applying advanced machine learning models and computational techniques, the study aims to predict and optimize the configuration of defects within TMDCs to achieve the desired performance characteristics. The successful realization of this objective will enable the rational design of materials with tailored properties, thereby accelerating the discovery and development of next-generation materials for various technological applications.

This paper explores the engineering of point defects in molybdenum disulfide (MoS2), a representative TMDC, to tailor its material properties using large language models (LLMs). By manipulating point defects, we aim to enhance the functional properties of MoS2 for specific technological applications, leveraging advanced computational tools to guide experimental efforts and optimize material performance.

\section{Related Work}

    One of the first works that explores property prediction with LLMs is \cite{fang2023mol}. The authors introduce a novel dataset which contains instructions in natural language for molecule and protein design. This dataset aims to enhance models understanding of biolmolecular features. Approaches to molecular property prediction with LLMs are introduced in \cite{liu2024moleculargpt}. \textit{MolecularGPT}, a fine-tuned language model that is trained on over 1000 property prediction tasks, is presented in this paper. The scope of the tasks the model can perform accurately is claimed to be narrowed down by zero- and few-shot prompting. 

Since crystal structures can be represented as graphs with a minimum unit cell repeating itself on a regular lattice in 3D space, Graph Neural Networks (GNNs) were studied for property prediction previously \cite{yan2022periodic}. However, the complexities of representing crystals structure nuances and incorporating important information, such as bond angles \cite{choudhary2021atomistic}, have hindered the practical application GNNs in the property prediction task.

GNNs are found to be overperformed by LLMs in \cite{rubungo2023llm} in the task if physical and electronic property prediction of crystals. \textit{LLM-Prop}, a method based on language models, is presented in this work as one that is able to predict the band gap, classify whether it as direct or indirect, and predicting unit cell volume. 

Inorganic compounds, such as crystals, can be described using the Crystallographic Information File (CIF) format. This text contains the information on structural properties of the crystals. The work \cite{antunes2023crystal} explores prediction of new crystals by fine-tuning LLMs on CIF format data. It is found that \textit{CrystaLLM}, a novel method for generating crystal structures, is capable of producing new valid crystal structures.

High public interest in natural language processing has pushed the progress in LLMs, including the open distribution of the largest models weights. The models with billions of parameters potentially contain most of the knowledge available online that can be leveraged to make further discoveries. The capabilities of fine-tuned Llama-2 with 70B parameters are explored in \cite{gruver2024fine}. The authors fine-tune this LLM on the tasks of generation of stable materials and infilling of partial structures. The stability rate of the generated materials is reported to be doubled in comparison with a diffusion model.

\section{Methods}
\subsection{2DMD, A 2D Material Defect Dataset}
In this study, we utilize a comprehensive dataset of 2D material defects to train and evaluate machine learning models aimed at predicting structure-property relationships. The dataset, referred to as the 2D Material Defect Database (2DMD) \cite{huang2023unveiling}, is designed to support a wide range of tasks, including the development of predictive models for electronic, optical, and mechanical properties of 2D materials. By providing a robust foundation for machine learning training, the dataset facilitates the exploration and optimization of material properties tailored for specific applications. 

The 2DMD dataset includes defect information for several widely studied 2D materials, namely molybdenum disulfide (MoS2), tungsten diselenide (WSe2), hexagonal boron nitride (h-BN), gallium selenide (GaSe), indium selenide (InSe), and black phosphorus (BP). These materials were selected due to their significant potential in electronic, optoelectronic, and flexible device applications. By focusing on these key materials, the dataset addresses a broad spectrum of material properties and defect configurations.

\begin{figure*}
    \centering
    \includegraphics[width=1\linewidth]{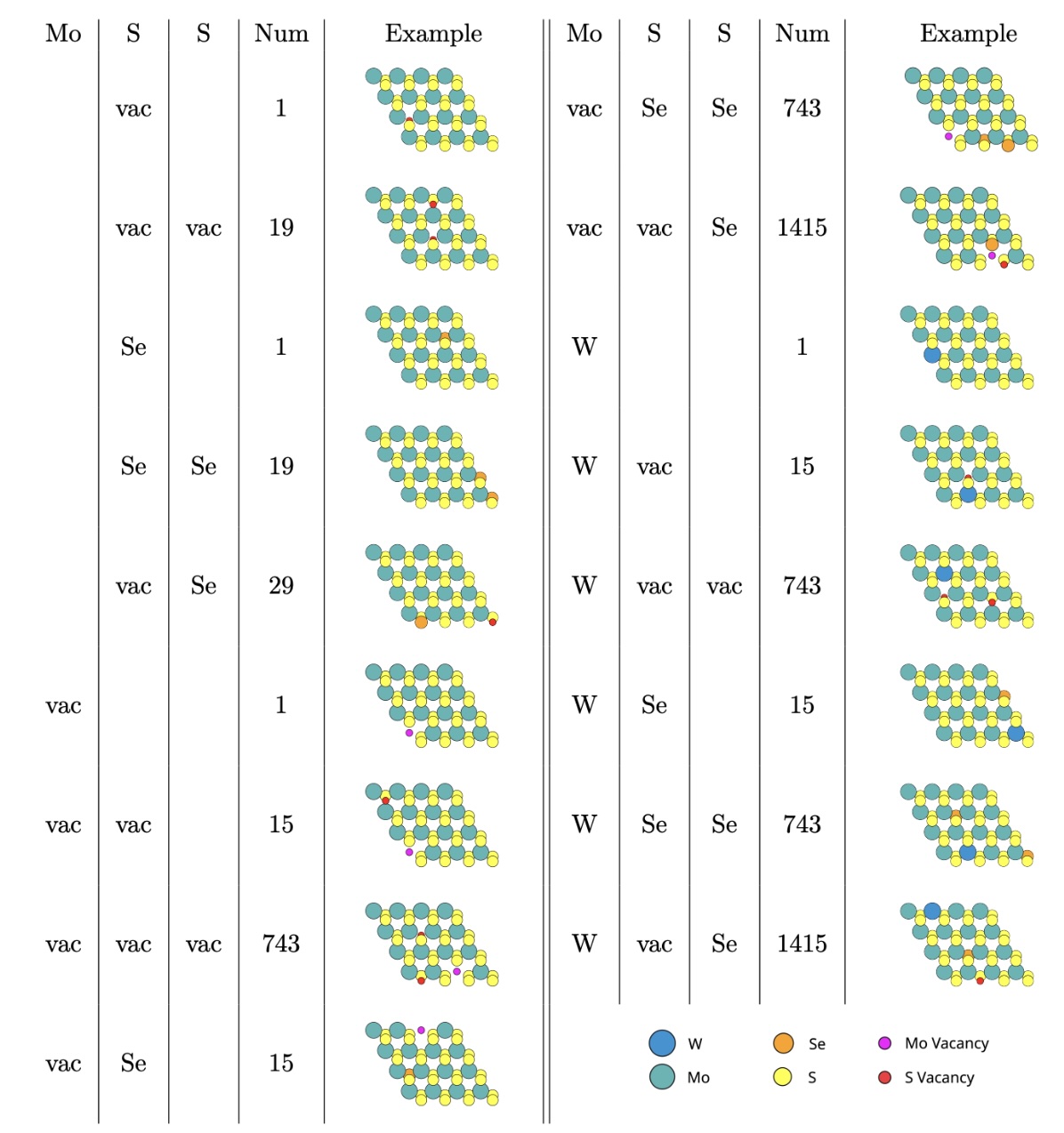}
    \caption{Defect types in low density dataset. The ‘Mo’ and two ‘S’ columns denote the the type of site that is being perturbed either by substituting the listed element, or a vacancy (vac). ‘Num’ column contains the number of structures with defects of the type in the dataset. Finally, ‘Example’ column presents a structure with such defect. Figure is taken from our previous work \cite{kazeev2023sparse}.}
    \label{fig:dataset}
\end{figure*}

The dataset is divided into two primary parts, each representing different defect concentration scenarios: low defect concentration and high defect concentration. The low defect concentration dataset comprises structured configurations, focusing on systematic defect placement within the crystal lattice. This part of the dataset includes 5933 structures each for MoS2 and WSe2, utilizing an 8x8 supercell configuration. The low defect concentration structures are meticulously crafted to explore the impact of isolated defects on material properties. These configurations provide insights into how single or few defects can alter electronic, optical, and mechanical behaviors, serving as a critical resource for understanding defect-material interactions. The defect types in the low density part of the dataset is described by Figure~\ref{fig:dataset}.

In contrast, the high defect concentration dataset consists of randomly generated configurations, featuring a higher density of defects within the crystal lattice. These configurations include both substitution and vacancy defects, reflecting more complex and disordered systems. By examining these structures, researchers can investigate the collective effects of multiple defects and the resulting modifications in material properties. This part of the dataset enables the study of defect interactions and their influence on material performance under more realistic and challenging conditions.

\begin{figure}
    \centering
    \includegraphics[width=0.9\linewidth]{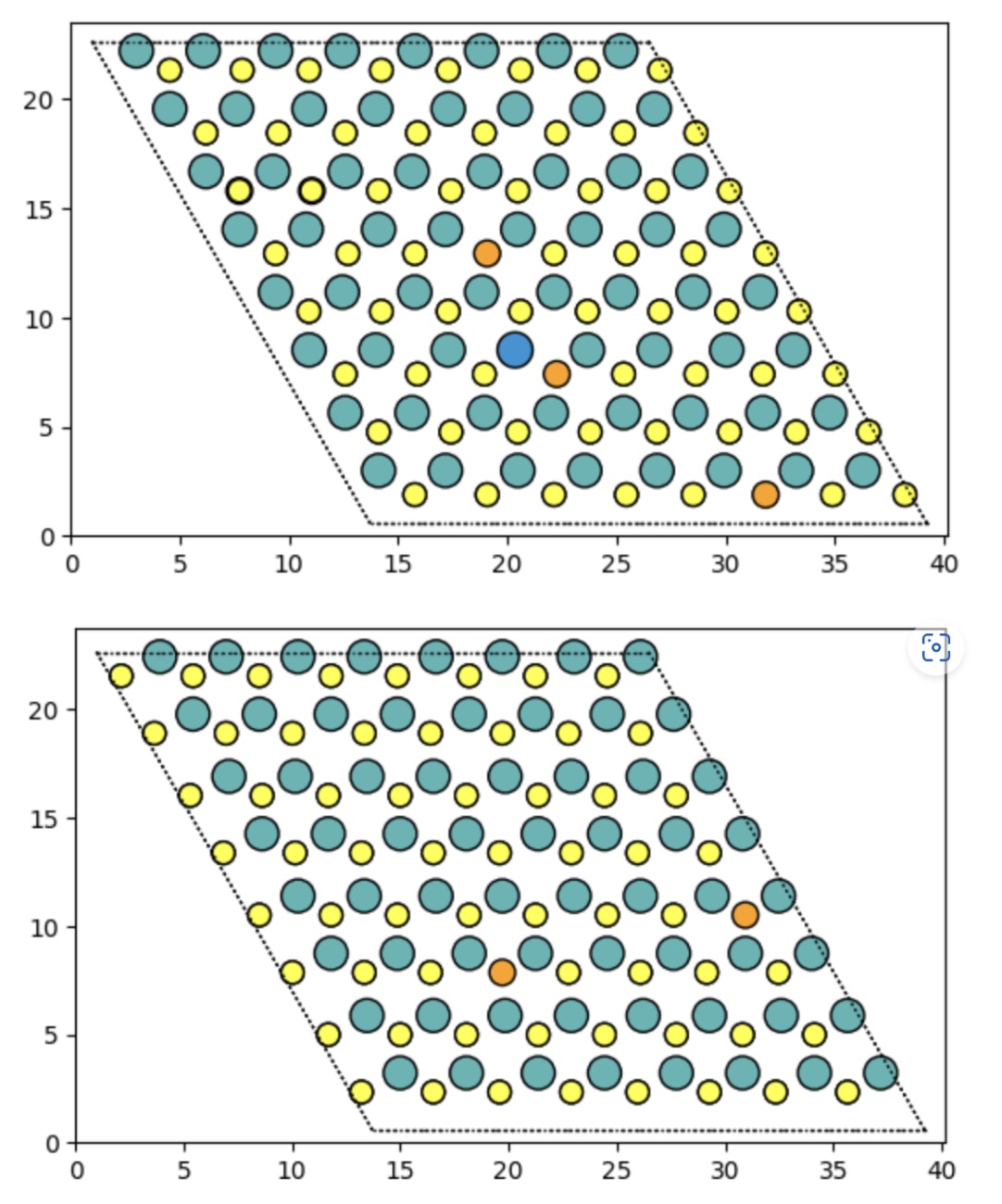}
    \caption{MoS2 crystal containing different defect configurations. The orange atoms refer to sulfur atoms substituted by selenium. The blue atoms are molybdenum atoms substituted with tungsten.}
    \label{fig:defects}
\end{figure}

Overall, the 2DMD dataset contains 14,866 distinct structures, each composed of 120 to 192 atoms. This extensive collection captures a wide variety of defect types, configurations, and concentrations, offering a comprehensive view of defect effects in 2D materials. The dataset is meticulously curated to ensure diversity and representation of different defect scenarios, providing a valuable resource for training advanced machine learning models. A sample of defects from the dataset can be found in Figure~\ref{fig:defects}.

The 2DMD dataset is instrumental for training machine learning models capable of predicting structure-property relationships in 2D materials. By leveraging this dataset, models can be developed to forecast key properties such as bandgap, conductivity, photoluminescence, and mechanical strength. The insights gained from these predictions can guide experimental efforts and accelerate the discovery and design of 2D materials with tailored properties for specific technological applications.

Furthermore, the dataset allows for the evaluation of machine learning methods, allowing researchers to benchmark and improve algorithms for defect property prediction. The comprehensive nature of the dataset, combined with its focus on diverse defect configurations, makes it an invaluable tool for advancing the field of 2D material research and the development of innovative materials for future technologies.

\subsection{Methodology}
The approach utilized in this study integrates advanced machine learning techniques with the Llama-3 large language model \cite{Llama3modelcard} to predict and generate crystal structures of transition metal dichalcogenides (TMDCs) with tailored defects. By leveraging the capabilities of large language models, this method aims to efficiently explore the vast configuration space of defects and optimize material properties for specific applications. We employ the largest Llama-3 available at the time of writing with 70B parameters. This choice is driven by the fact that this model has scored the highest out of all other Llama models on natural language benchmarks \cite{Llama3modelcard}.

A parameter-efficient fine-tuning method, known as Low-Rank Adaptation of Large Language Models (LoRA) \cite{hu2021lora}, was employed to adapt the Llama model to the specific task of crystal structure generation. LoRA enables efficient fine-tuning by injecting rank decomposition matrices into every transformer layer and reducing the number of trainable parameters, thereby allowing the model to learn relevant features from a subset of the Crystallographic Information File (CIF) data. This fine-tuning process is critical for aligning the model's capabilities with the domain-specific requirements of TMDC defect engineering. Table~\ref{tab:hyperparams_ft} describes the fine-tuning hyperparameter setup.

\begin{figure}
    \begin{center}
    \mdfsetup{%
        middlelinecolor=black,
        middlelinewidth=1pt,
        backgroundcolor=gray!10,
        roundcorner=5pt,
        frametitle={Instruction Prompt},
        frametitlerule=true}
    \begin{mdframed}
    \texttt{Here is a TMDC material (MoS2), unitcell of size [8, 8, 1].\\
    The defects for this material BN are generated based on following rules:\\
    \{'type': 'substitution', 'from': 'S', 'to': 'Se'\}\\
    \{'type': 'substitution', 'from': 'Mo', 'to': 'W'\}\\
    \{'type': 'vacancy', 'element': 'Mo'\}\\
    \{'type': 'vacancy', 'element': 'S'\}.\\
    The material has the following properties:\\
    - The formation energy per atom is 6.1508.\\
    - The energy per atom is -8.6699.\\
    - The Fermi level is -4.1761.\\
    Generate defects in the crystal structure, of the following types (vacancy, substitutional).\\
    You are only allowed to change the atom symbol (in the case of vacancy replace it with VAC) without changing anything else.\\
    Crystal structure:\\
    25.5 25.5 14.9\\
    90 90 119\\
    Mo\\
    0.71 0.39 0.49\\
    Mo\\
    0.71 0.51 0.49\\
    Mo\\
    0.71 0.64 0.49\\
    Mo\\
    0.71 0.76 0.49\\
    Mo\\
    0.83 0.76 0.49\\
    Mo\\
    0.83 0.89 0.49…\\
    <Crystal elements and coords>}
    \end{mdframed}
    \caption{A sample of instruction for LLM fine-tuning.}
    \label{fig:instructions}
    \end{center}
\end{figure}

To facilitate the training of the Llama model, we construct an instruction dataset based on the 2D Material Defect Database (2DMD). An instruction sample can be found in Figure~\ref{fig:instructions}. This dataset includes detailed descriptions of crystal structures, defect substitution rules, and associated properties. The dataset captures various defect types and configurations, along with their restrictions and impact on material properties. These instructions serve as the foundation for generating new crystal structures with desired attributes.

\begin{table}%[htbp]
\caption{Fine-tuning hyperparameters.}
\begin{tabular}{c|c}\hline
\textbf{Parameter} & \textbf{Value} \\ \hline
Learning rate & $1e-4$ \\ \hline
Learning rate scheduler & Cosine \\ \hline
Number of epochs & 10 \\ \hline
LoRA rank & 8 \\ \hline
LoRA dropout & 0.05 \\ \hline
\end{tabular}\label{tab:hyperparams_ft}
\end{table}

To enhance the robustness and generalizability of the model, data augmentation techniques were applied. Random rotations are used to generate diverse orientations of crystal structures, simulating real-world variations and increasing the model's exposure to different configurations. Additionally, physical properties were randomly sampled to provide a wide range of target attributes for model training, allowing the model to learn associations between structural features and material properties.

The inference process involves using the fine-tuned Llama model to generate new crystal structures based on input instructions and masked coordinates. The model receives detailed instructions specifying the types of defects, any restrictions on their configurations, and target physical properties. It then generates the corresponding coordinates of the crystal structure, which are parsed to create a viable model of the material. We run the inference through the VLLM framework \cite{kwon2023efficient} to speed up the generation. The inference hyperparameters are listed in the Table~\ref{tab:hyperparams_inf}.

\begin{table}%[htbp]
\caption{Inference hyperparameters.}
\begin{tabular}{c|c}\hline
\textbf{Parameter} & \textbf{Value} \\ \hline
Temperature & 1.0 \\ \hline
\texttt{Top\_p} & 0.9 \\ \hline
LoRA alpha & 32 \\ \hline
\end{tabular}\label{tab:hyperparams_inf}
\end{table}

Once the coordinates are generated, the energies of the resulting structures are calculated using a surrogate model \ref{sec:surrogate_mode}. This step involves evaluating the stability and feasibility of the predicted structures, ensuring that the generated configurations meet the desired energy criteria and other specified properties. The use of a surrogate model allows for efficient energy calculations, facilitating rapid assessment of a large number of potential structures.

\subsection{Property prediction} \label{sec:surrogate_mode}
The properties of the generated crystals in this study have been calculated using a surrogate model that integrates a simplified line-input crystal-encoding system (SLICES) \cite{xiao2023invertible} with gradient boosted decision trees. This approach leverages the principles of quotient graphs and Eon's method \cite{eon2011euclidian} for reconstructing crystal structures, specifically utilizing Euclidean embeddings of periodic nets to provide a topologically induced complete set of geometric descriptors for crystal structures.

% Simplified Line-Input Crystal-Encoding System (SLICES)
The SLICES methodology is designed to efficiently encode crystal structures by capturing their essential periodic characteristics. It uses quotient graphs to represent the repeating units and their connectivity within the crystal lattice. This allows for a compact and precise representation of periodic crystals, which is crucial for handling the inherent complexity of these materials.

Quotient graphs are particularly effective in this context because they reduce the infinite repetition of periodic crystals to a finite and manageable form. By focusing on the fundamental repeating units, quotient graphs capture the essence of the crystal's topology and symmetry. Eon's method complements this by reconstructing these structures into Euclidean space, ensuring that the generated embeddings maintain their geometric and topological integrity.

% Decision Trees and SLICES Integration
To predict the properties of the encoded crystal structures, Catboost \cite{dorogush2018catboost}, an implementation of gradient boosted decision trees, is employed. Catboost is chosen for its ability to model complex non-linear relationships and handle heterogeneous data efficiently. It excels in scenarios where categorical features are present and is robust against overfitting, making it suitable for predicting the diverse properties of defect engineered crystals.

Gradient boosted decision trees work by creating an ensemble of decision trees, where each tree is trained to correct the errors of its predecessor. This iterative process leads to a model that can capture intricate patterns in the data, resulting in accurate predictions. In this study, the SLICES crystal line strings, which encapsulate the encoded structure of crystals, are used as inputs for the Catboost model. The model predicts various physical properties, including energy states, by learning from the encoded data .

% Property Prediction Process
The process begins with the generation of a crystal structure, which is then encoded into a line string using the SLICES method. This encoding captures the key structural features and periodicity of the crystal. The encoded line string is subsequently input into the Catboost model, which predicts the desired properties. These properties include electronic, optical, and mechanical characteristics, which are critical for evaluating the potential applications of the generated crystals.

The integration of SLICES and Catboost offers several advantages. It allows for the efficient exploration of the vast chemical space associated with defect-engineered crystals, facilitating the discovery of materials with tailored properties. Furthermore, the approach is scalable and can be applied to a wide range of crystal structures, making it a versatile tool for materials science research.

In summary, this surrogate model, combining SLICES and Catboost, provides a powerful framework for predicting the properties of generated crystals. By leveraging the strengths of quotient graphs and machine learning, the model enables accurate and efficient evaluation of crystal properties, paving the way for the development of advanced materials with engineered defects .

\section{Results}

The surrogate model is used to evaluate the quality of the dataset comprising 6000 synthetically generated MoS2 crystals with varied defect concentrations 
% The evaluation and results of the surrogate model applied to a dataset comprising 6000 synthetically generated MoS2 crystals with varied defect concentrations. 
The surrogate model is tasked with predicting several crystal properties, and its performance was gauged using refined statistical metrics focused on the top-performing predictions.

The evaluation of the surrogate model's predictions is confined to the top 25\% of predictions with the lowest mean squared error (MSE), in other words structures with large error values are ignored due to the low fidelity of our surrogate model. The performance metrics used are $R^2$ and Root Mean Square Error (RMSE), calculated as follows:

$R^2$ (Coefficient of Determination):
   \[ R^2_{25th} = 1 - \frac{\sum_{i \in I}{(y_i - \hat{y}_i)^2}_{\{i \mid (y_i - \hat{y}_i)^2 \leq Q_{0.25}\}}}{\sum_{i \in I}{(y_i - y_{25th})^2}_{\{i \mid (y_i - \hat{y}_i)^2 \leq Q_{0.25}\}}} 
   \]
   Where \( y_i \) is the observed value, \( \hat{y}_i \) is the predicted value, and \( y_{25th} \) is the $25^{th}$ percentile of observed values. This metric reflects the proportion of the variance in the dependent variable that is predictable from the independent variables, specifically focusing on the subset of data with the most accurate predictions.

RMSE (Root Mean Square Error):
   \[ RMSE_{25th} = \sqrt{\frac{1}{|I|} \sum_{i \in I}{(y_i - \hat{y}_i)^2}_{\{i \mid (y_i - \hat{y}_i)^2 \leq Q_{0.25}\}}} \]
   This metric provides a measure of the average magnitude of the prediction errors, scaled specifically for the best-performing subset of the data.

% \subsection{Discussion}
The surrogate model demonstrated robust predictive performance across various crystal properties. For instance, the Formation Energy Per Site exhibited an $R^2$ of 0.99 and an RMSE of 0.07, indicating near-perfect predictive accuracy within the most reliable quartile of predictions. Similar high-performance metrics were observed for other properties, see table \ref{table:results}.
\begin{table}[ht]%[htbp]
\label{table:results}
\caption{Property quality estimation of the generated defects configurations using surrogate model}
\begin{tabular}{c|c|c}\hline
\textbf{Property} & \textbf{$R^2$} & \textbf{RMSE}\\ \hline
Homo-Lumo Gap & 0.98 & 0.05 \\
Homo & 0.95 & 0.06 \\
Lumo & 0.82 & 0.06 \\
Formation Energy Per Site & 0.99 & 0.07 \\
\end{tabular}\label{tab:results}
\end{table}

\begin{figure*}
    \centering
    \includegraphics[width=1\linewidth]{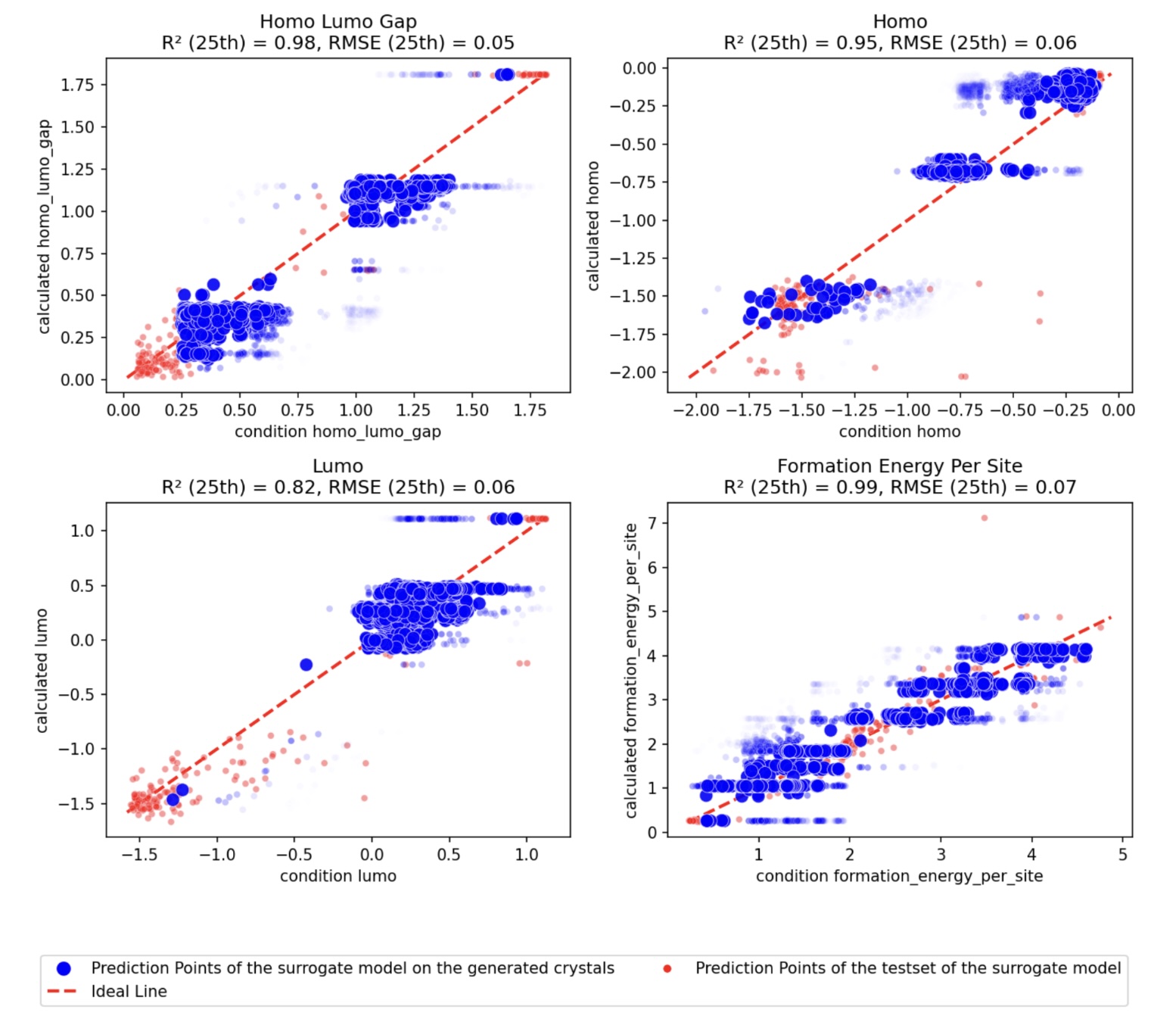}
    \caption{6000 generated MoS2 crystals containing different defect concentrations evaluated with surrogate model. Structures with high MSE from the target property have reduced opacity and size.}
    \label{fig:results}
\end{figure*}

% Visualization Insights
The scatter plots (Figure ~\ref{fig:results}) visually articulate the distribution and accuracy of predictions. Points are color-coded to distinguish between predictions on generated crystals (blue) and those on the test set (red). Structures with higher MSE have been represented with reduced opacity and size, which visually segregates them from more accurate predictions, thus allowing for a quick visual assessment of model performance spread.

This detailed statistical and visual analysis confirms the efficacy of the surrogate model in accurately predicting properties of MoS2 crystals, particularly emphasizing its reliability for the best-predicted subset of data. The model not only serves as a tool for rapid screening of crystal properties but also highlights the potential for predictive models in guiding synthetic strategies and understanding material behavior under varying conditions.

\section{Discussion}
The decision to employ LoRA (Low-Rank Adaptation) instead of full fine-tuning for adapting large language models (LLMs) to our specific task of predicting crystal properties was driven primarily by considerations of computational efficiency. LoRA allows for modifying only a small portion of the model's weights—specifically, low-rank matrices—thus significantly reducing the computational resources required for training and inference. This approach maintains the pre-trained knowledge of the LLM while allowing sufficient flexibility to learn taskspecific nuances, thereby offering a balance between performance and efficiency.

Large language models, particularly those fine-tuned for specialized domains, have shown remarkable capacity to categorize and predict based on structured data, such as defect configurations in materials science. Our model leverages the ability to classify potential defects into a limited number of defect classes. Since these classes can be topologically equivalent, the model learns to recognize these equivalences and map each class to specific physical attributes. This capability is akin to how LLMs manage language: by recognizing patterns and mapping them to meanings, albeit here the "meanings" are physical properties linked to specific structural defects.

The decision not to train directly on Simplified Line-Input Crystal-Encoding System (SLICES) and instead use Crystallographic Information File (CIF) formats was informed by the limitations of the existing algorithms within the scope of Eon's theory on Euclidian embeddings of periodic nets. Eon's method is designed for fully connected crystal graphs with a translation vector rank of three or higher. Given that transition metal dichalcogenides (TMDCs) like MoS2 are two dimensional materials with a translation vector rank of two, they fall outside the applicability of Eon’s embedding method. This limitation is critical as the reconstruction algorithm of SLICES, based on these embeddings, fails to support low-dimensional materials, necessitating the use of CIF, which does not rely on these embeddings.

The choice to utilize the Llama-3 model for this task was guided by several considerations. Firstly, Llama models are at the forefront of performance among open-source large language models, ensuring state of the art results. Additionally, the Llama-3 model's approach to tokenizing numbers treating digits individually provides a more granular and, consequently, potentially more precise handling of numerical data common in material science datasets. This characteristic makes Llama particularly suitable for tasks involving detailed quantitative predictions, such as those required for accurately mapping defect types to their physical impacts in crystal structures.

These methodological choices collectively enable the effective application of advanced AI techniques to the domain of material science, specifically in the modeling and prediction of physical properties based on structural defects in crystals. The adoption of these advanced computational strategies ensures both efficiency and high fidelity in predictions, essential for advancing materials design and characterization.

\begin{acknowledgments}
The article was prepared within the framework of the project ``Mirror Laboratories'' HSE University, RF.
\end{acknowledgments}

\newpage
%%%%%%%%%%%%%%%%%%%%%%%%%%%%%%%%
% USE thebibliography
%%%%%%%%%%%%%%%%%%%%%%%%%%%%%%%%

% \bibliography{bib}

\end{document}